\documentclass[aps,prd,preprint,nofootinbib]{revtex4}
\usepackage{amsmath,amssymb,graphics,epsfig,multirow}%,mathbbol,graphicx,psfrag}
\newcommand{\newc}{\newcommand}
\newc{\bsym}{\boldsymbol}
\newc{\mrm}{\mathrm}
\newc{\ovl}{\overline}
\newc{\ovla}{\overleftarrow}
\newc{\ovra}{\overrightarrow}
\newc{\ra}{\rightarrow}
\newc{\lra}{\leftrightarrow}
\newc{\wtil}{\widetilde}
\newc{\eps}{\epsilon}
\newc{\hc}{\dagger}
\newc{\pd}{\partial}
\newc{\SL}{\!\!\!/}
\newc{\LH}{\hat{L}}
\newc{\RH}{\hat{R}}
\newc{\sWsq}{\sin^2\theta_\mathrm{W}}
\newc{\cWsq}{\cos^2\theta_\mathrm{W}}
\newc{\half}{\frac{1}{2}}

\begin{document}
\title{Discrete Dirac Neutrino in Warped Extra Dimensions}
\author{We-Fu Chang}
\email{wfchang@phys.nthu.edu.tw}
\affiliation{Department of Physics, National Tsing Hua University, Hsin Chu 300, Taiwan}
\author{John N. Ng}
\email{misery@triumf.ca}
\affiliation{Theory Group, TRIUMF, 4004 Wesbrook Mall, Vancouver, B.C., Canada}
\author{Jackson M. S. Wu}
\email{jbnwu@itp.unibe.ch}
\affiliation{Institute for Theoretical Physics, University of Bern, Sidlerstrasse 5, 3012 Bern, Switzerland}

\date{\today}

\begin{abstract}
We implement Dirac neutrinos in the minimal custodial Randall-Sundrum setting via the Krauss-Wilczek mechanism. We demonstrate by giving explicit lepton mass matrices that with neutrinos in the normal hierarchy, lepton mass and mixing patterns can be naturally reproduced at the scale set by the constraints from electroweak precision measurements, and at the same time without violating bounds set by lepton flavour violations. Our scenario generically predicts a nonzero neutrino mixing angle $\theta_{13}$, as well as the existence of sub-TeV right-handed Kaluza-Klein neutrinos, which partner the right-handed Standard Model charged leptons. These relatively light KK neutrinos may be searched for at the LHC.
\end{abstract}


\maketitle

\section{Introduction}
It is now generally accepted that neutrinos have small masses, and a phenomenology of neutrino oscillations
within the framework of massive neutrino mixings is by now well developed. However, the nature of the neutrino -- whether it is a Dirac or Majorana particle -- and thus the origin of neutrino masses remains unknown. Much of the vast literature on the origin of neutrino masses invokes the see-saw mechanism, and so focuses on models of Majorana neutrinos. Since left-handed (LH) leptons in the SM are charged under
$SU(2)_L \times U(1)_Y$, only singlet right-handed (RH) neutrinos can take on tree level Majorana masses. To forbid such mass terms, RH neutrinos are required to carry internal charges, and these are usually assumed to arise from a $U(1)$ symmetry -- be it global or gauged -- that is commonly identify as the lepton number. A massive Dirac neutrino can then arise if LH leptons carry the appropriate charges that allow Yukawa couplings to RH neutrinos and the Higgs boson. Note that the internal symmetry need not be a $U(1)$ symmetry, although it is usually implicitly assumed. Indeed, it has long been known that discrete symmetries can lead to Dirac neutrinos. These models are typically constructed within the context of supersymmetric models, where the low-scale discrete symmetries come as remnants of some broken high-scale gauge symmetry, and are to be understood as gauged discrete symmetries~\cite{KW88,B89,PK90}.

In this paper we study the viability of Dirac neutrinos in the context of warped Randall-Sundrum (RS) extra-dimensional scenario~\cite{RS}. This is a departure from previous studies, which aim at incorporating the see-saw mechanism in the RS scenario~\footnote{A see-saw model with an ``almost'' Dirac neutrino was constructed in~\cite{Gher03}.}. Recent studies have shown that the RS scenario provides a novel and powerful framework to understand flavour physics (see e.g.~\cite{APS05} and references within). The crucial point is that the observed SM charged fermion mass hierarchy can be naturally produced from their ``geography'' in the five-dimensional (5D) $AdS_5$ bulk~\cite{FermLoc}, which also solves the electroweak hierarchy problem~\cite{RS}. The fermion mass hierarchy now arises from the overlap of fermion wavefunctions in the bulk, whose form is determined by the $AdS_5$ geometry, and whose location given by the fermion bulk mass (or localization) parameters. Neither these nor the Yukawa couplings need be fine-tuned, and Yukawa couplings can be naturally of order one with a completely random pattern, i.e. ``anarchic''~\cite{HS01,APS05}. Indeed, from this approach, the observed quark mass and mixing patterns can be accurately reproduced~\cite{CNW08}. Thus, it is reasonable to expect that Dirac neutrinos can have naturally small masses without fine-tuning from the appropriate localization of neutrinos in the bulk, which we find is indeed the case. This is in sharp contrast to the usual 4D Dirac neutrino models where excessive fine-tuning is often required.

For the framework of our study, we choose the minimal custodial RS (MCRS) model first given in~\cite{CusRS}. The MCRS model has $SU(2)_L \times SU(2)_R \times U(1)_{B-L}$ as its bulk gauge group, which encodes a custodial $SU(2)$ symmetry that protects the $\rho$ parameter from excessive corrections due to Kaluza-Klein (KK) excitations of the bulk fields. Matter fields reside in the bulk, and the SM chiral fermions are idendified as the zero-modes of bulk fermion fields. This set-up is fully realistic and satisfies all constraints from precision electroweak precision tests (EWPTs). To have Dirac type active neutrinos, we forbid Majorana mass terms for the RH neutrino zero-modes by a discrete $Z_N$ symmetry obtained via the Krauss-Wilczek (KW) mechanism~\cite{KW88} from a gauged $U(1)$ symmetry in the bulk. In this paper, we will focus on the simplest case where we augment the MCRS bulk gauge group by an additional $U(1)_X$~\footnote{An extra $U(1)$ can also be used to suppress proton decay. A mechanism for doing so is discussed in~\cite{ADS03}.}. The KW mechanism can then be implemented straightforwardly with the help of UV localized boundary Higgs.

As small neutrino masses can be obtained naturally through neutrino geography, the challenge then is to accommodate the large neutrino mixings, which is very different from quark mixings. One would also like to achieve all these at a scale relevant for the LHC. For the MCRS model, EWPTs set the scale of KK resonances at around 3~TeV~\cite{CusRS}. However, flavour changing neutral current (FCNC) processes resulting from the anarchic 5D flavour structure generally push the KK scale up beyond the reach of the LHC, particularly in the quark sector where FCNC constraints are severe~\cite{CFA08}. To bring the KK scale down to a few TeV order, additional flavour symmetries have been proposed (see e.g.~\cite{CFA09}). In the lepton sector, which we concentrate on in this paper, the flavour constraints on the KK scale is less severe. With some tuning in the Yukawa couplings, we find that for the normal hierarchy, the observed neutrino mass spectrum and mixing pattern can be reproduced very well within the MCRS set-up with the KK scale at the 3~TeV level; fitting for the case of inverted hierarchy or degenerate neutrinos requires either excessive fine-tuning or a much higher KK scale, or both. Alternative approaches include imposing an additional lepton flavour symmetry (see e.g.~\cite{CDGG08}) or lepton minimal flavour violation (see e.g.~\cite{PR09}). An interesting fact from our set-up is that there are generically RH KK neutrinos with mass in the range of $\mathcal{O}(10 - 100)$~GeV. This follows from the localization of charged leptons necessary to reproduce charged lepton masses. These moderately heavy sub-TeV KK neutrinos will be particularly interesting for the LHC to search for.

The paper is organized as follows. In Sec.~\ref{sec:Setup}, we describe our set-up for dirac neutrinos. We first briefly describe aspects of the MCRS model relevant for our study to set up notations. We then show how the KW mechanism can be implemented to generate a gauged discrete $Z_N$ symmetry in the 4D effective theory  that forbids Majorana mass terms. In Sec.~\ref{sec:Configs}, we scan the parameter space for lepton configurations that are consistent with the current charged lepton and neutrino data. We give five representative viable configurations for the case of neutrino normal hierarchy. In Sec.~\ref{sec:Pheno}, we study the phenomenology of the relatively light RH KK neutrinos, which can give interesting signals observable at the LHC. Sec.~\ref{sec:Conc} contains our conclusions.

\section{\label{sec:Setup} The set-up}
\subsection{\label{subsec:MCRS} Fermions in the MCRS model}
The MCRS model is formulated on a slice of $AdS_5$ space with the fifth dimension compactified on an
$S_1/(Z_2 \times Z_2')$ orbifold. The background metric given by
\begin{equation}\label{Eq:metric}
ds^2 = G_{AB}\,dx^A dx^B = e^{-2\sigma(\phi)}\,\eta_{\mu\nu}dx^{\mu}dx^{\nu}-r_c^2 d\phi^2 \,, \quad
\sigma(\phi) = k r_c |\phi| \,,
\end{equation}
where $\eta_{\mu\nu} = \mathrm{diag}(1,-1,-1,-1)$, $-\pi\leq\phi\leq\pi$, $k$ the $AdS_5$ curvature and $r_c$ the compactification radius. A UV (Planck) brane sits at the orbifold fixed point $\phi=0$, and an IR (TeV) brane at $\phi=\pi$. To solve the hierarchy problem, we take $k r_c\pi \approx 37$. A warped down scale defined by $\tilde{k} = k e^{-k r_c\pi}$ sets the scale of the first KK gauge boson mass,
$M_{KK} \approx 2.45\tilde{k}$, and thus the scale of the new physics.

The SM electroweak gauge group is extended to $SU(2)_L \times SU(2)_R \times U(1)_{B-L}$ in the bulk to incorporate the custodial symmetry, which is reduced on the boundary branes. Localized on the IR brane, the SM Higgs, $H_1$, now transforms as a bidoublet under $SU(2)_L \times SU(2)_R$, which breaks down to $SU(2)_D$ when $H_1$ acquires a vacuum expectation value (VEV). On the UV brane, boundary conditions break
$SU(2)_R \times U(1)_{B-L}$ down to $U(1)_Y$.

The SM fermions are all embedded as doublets in the bulk through the use of 5D Dirac spinors. In particular, there is a separate doublet for every SM lepton, while we choose to embed RH neutrinos as singlets:
\begin{align}\label{Eq:leprep}
L_i &=
\begin{pmatrix}
\nu_{iL}\,[+,+] \\
e_{iL}\,[+,+]
\end{pmatrix} \,,\quad
E_i =
\begin{pmatrix}
\tilde{\nu}_{iR}\,[-,+] \\
e_{iR}\,[+,+]
\end{pmatrix} \,, \quad
\nu_{iR}\,[+,+]  \,,
\end{align}
where $i$ is a generation index, $L$ ($E$) denotes $SU(2)_L$ ($SU(2)_R$) doublet for the LH (RH) charged leptons, and $\nu_{R}$ denotes the RH neutrinos singlet under both $SU(2)_L$ and $SU(2)_R$. The parity assignment $+$ ($-$) denote Neumann (Dirichlet) boundary conditions (BCs) applied to the spinors on the boundary branes. Only fields with the [+,+] parity contain zero-modes that are part of the low energy spectrum of the 4D effective theory.

A given 5D bulk fermion, $\Psi$, can be KK expanded as
\begin{equation}\label{Eq:PsiKK}
\Psi_{L,R}(x,\phi) = \frac{e^{3\sigma/2}}{\sqrt{r_c\pi}}
\sum_{n=0}^\infty\psi^{(n)}_{L,R}(x)f^n_{L,R}(\phi) \,,
\end{equation}
where subscripts $L$ and $R$ label the chirality, and the KK mode wavefunction $f^n_{L,R}$ are normalized according to
\begin{equation}\label{Eq:fnorm}
\frac{1}{\pi}\int^\pi_{0}\!d\phi\,f^{n*}_{L,R}(\phi)f^m_{L,R}(\phi) = \delta_{mn} \,.
\end{equation}
The KK wavefunctions are obtained from solving the equations of motion. In particular, the zero-mode wavefunctions are given by
\begin{equation}
f^0_{L,R}(\phi,c_{L,R}) =
\sqrt{\frac{k r_c\pi(1 \mp 2c_{L,R})}{e^{k r_c\pi(1 \mp 2c_{L,R})}-1}}
e^{(1/2 \mp c_{L,R})k r_c\phi} \,,
\end{equation}
where $c_{L,R}$ are the bulk Dirac mass parameters defined by $m = c\,k$, and the upper (lower) sign applies to the LH (RH) label. Depending on the orbifold parity of the fermion, one of the chiralities is projected out.

After KK reduction, couplings of KK modes in the 4D effective theory arise from the overlap of the wave
functions in the bulk. For the interaction between a {\it q}th KK gauge boson and an {\it m}th and an 
{\it n}th KK fermion, the coupling is given by
\begin{equation}\label{Eq:gffA}
g^{m\,n\,q}_{f\bar{f}A} =
\frac{g_4}{\pi}\int^\pi_{0}\!d\phi\,f^m_{L,R}(\phi)f^{n*}_{L,R}(\phi)\chi_q(\phi)\,,
\end{equation}
where $g_4 \equiv g_5/\sqrt{r_c\pi}$ is the 4D coupling.

The Yukawa interactions between the SM Higgs and SM charged fermions are localized on the IR brane. They lead to mass terms for SM fermions in the 4D effective theory after electroweak symmetry breaking (EWSB). To generate Dirac masses for the neutrinos, we introduce another Higgs on the IR brane, $H_2$, which only transforms nontrivially as a doublet under the $SU(2)_L$. Note that the $H_2$ behaves like a second Higgs doublet in the 4D extended Higgs model. Also since $H_2$ is an $SU(2)_R$ singlet, it does not couple to the SM charged fermions and so will not affect their phenomenology.

After EWSB, fermion masses in the 4D effective theory take the general form
\begin{align}
\int\!d^4x\,\frac{1}{k r_c\pi}\Big[
&v_1 \ovl{Q}(x,\pi)\lambda^u_{5}U(x,\pi) + v_1\ovl{Q}(x,\pi)\lambda^d_{5}D(x,\pi) + \notag \\
&v_1 \ovl{L}(x,\pi)\lambda^e_{5}E(x,\pi) + v_2 \ovl{L}(x,\pi)\lambda^\nu_{5}\nu_R(x,\pi)
+ \mrm{h.\,c.}\Big] \,,
\end{align}
where $v_1$ and $v_2$ are the VEVs of $H_1$ and $H_2$ respectively, and $\lambda_5^f$ denotes the complex dimensionless 5D Yukawa matrix for each fermion species $f$. For zero-modes, this gives the mass matrices for the SM fermions in the 4D effective theory
\begin{equation}\label{Eq:RSM}
(M^{RS}_f)_{ij} = \frac{v_f}{k r_c \pi}\lambda^f_{5,ij}f^0_{L}(\pi,c^{f}_{iL})f^0_{R}(\pi,c^{f}_{jR}) \,, \qquad f = u,\,d,\,e,\,\nu.
\end{equation}
where $v_{u,\,d,\,e}=v_1$ and $v_\nu=v_2$.  One expects that the two VEVs would not be too different, as both should be related to $\tilde{k}$, the natural scale in the 5D theory warped down. For simplicity, we take
$v_1 = v_2 = v_W/\sqrt{2}$ in our study, where the pattern of EWSB fixes $v_W = \sqrt{v_1^2+v_2^2}= 174$~GeV.

The mass matrices are diagonalized by a bi-unitary transformation
\begin{equation}
(U_L^f)^\hc M^{RS}_f\,U_R^f =
\begin{pmatrix}
m^f_1 & 0     & 0 \\
0     & m^f_2 & 0 \\
0     & 0     & m^f_3
\end{pmatrix} \,,
\end{equation}
where $m^f_i$ are the mass eigenvalues, and the mass eigenbasis is defined by $\psi' = U^\hc\psi$. Then for quarks, the CKM matrix is given by $V_{CKM} = (U^u_L)^\hc U^d_L$, while for leptons, the PMNS matrix is given by $V_{PMNS} = (U^e_L)^\hc U^\nu_L$.

\subsection{\label{sec:ZN} Gauged discrete $Z_N$ symmetry and Dirac neutrinos}
In order to have Dirac neutrinos, Majorana mass terms from $v_R$ have to be forbidden. A simple way to do this is to have an additional $U(1)$ symmetry. Because quantum gravity effects do not respect global symmetries~\cite{KW88}, this $U(1)$ has to be gauged. But gauged symmetry has to be broken, as otherwise new massless gauge boson would appear. Nevertheless, through KW mechanism a gauged discrete $Z_N$ symmetry can remain after breaking the $U(1)$ gauge symmetry on the UV brane (see below), and so Majorana neutrino mass terms stay forbidden.

To implement the KW mechanism, we extend the MCRS bulk gauge group by an additional $U(1)_X$. This is then broken spontaneously on the UV brane via a UV brane-localized scalar, $\phi$. The covariant derivative of $\phi$ is given by
\begin{equation}
D_\mu\phi = \left(\pd_\mu - i g_{5X} X_\mu\right)\phi \,,
\end{equation}
where $X_\mu$ is the $U(1)_X$ gauge field, and $g_{5X}$ the gauge coupling constant. After spontaneous symmetry breaking $\phi$ acquires a VEV, $v_\phi$, and it can be parametrized as
$\phi = (v_\phi + \rho)e^{i\eta/v_\phi}$. The Goldstone field, $\eta$, can be removed by a gauge transformation accompanied by a concomitant redefinition of the fermion field:
\begin{equation}
X_\mu \ra X_\mu - \frac{1}{g_{5X}}\frac{\pd_\mu\eta}{v_\phi} \,, \qquad
f \ra f\exp\left(i\frac{\eta}{v_\phi}Q_X\right) \,.
\end{equation}
The $Z_N$ symmetry then emerges from the $U(1)_X$ symmetry if $Q_X$, the fermion charge under the $U(1)_X$, is rational but nonintegral. As is reviewed in Appendix~\ref{app:4DDN}, the usual results for Dirac neutrinos in 4D -- where a $Z_N$ is put in by hand -- can be carried over, and the smallest group is $Z_3$. We assume that IR localized Higgs fields, $H_{1,2}$, are singlets under the $Z_N$, so the gauged discrete symmetry is exact in the 4D effective theory.

\subsection{Neutrinoless double beta decay}
Interestingly, the discrete charge also forbids neutrinoless double beta , $0\nu\beta\beta$, decays  in nuclei. The reason is that if the SM fermions carry any charges, $\alpha^f$, other than the SM gauge charges, $0\nu\beta\beta$ decays must satisfy the following condition:
\begin{equation}\label{eq:0nu}
\alpha^{d}_{\chi_1}+\alpha^{d}_{\chi_2} =
\alpha^{u}_{\chi_3}+\alpha^{u}_{\chi_4}+\alpha^{e}_{\chi_5}+\alpha^{e}_{\chi_6} \,,
\end{equation}
where $\chi_i$ labels the fermion chirality. This is trivially satisifed in models of Majorana neutrinos.
In our case of Dirac neutrinos from a discrete symmetry, we have $\alpha^{d}_\chi = \alpha^{u}_\chi$ independent of the chiralities of the quarks, and also $\alpha_{e_L}=\alpha_{e_R}\neq 0$ (see Eq.~\eqref{eq:charges}). Thus, Eq.~\eqref{eq:0nu} is not satified, and $0\nu\beta\beta$ decays are forbidden.

Note that assigning a lepton number for both LH and RH electrons using a $U(1)$ symmetry, as is commonly done, will also {\it not} satify Eq.~\eqref{eq:0nu}. As a consequence, $0\nu\beta\beta$ decays in nuclei cannot be used to determine whether Dirac neutrinos arise from a continuous symmetry, such as a $U(1)_L$ lepton number, or a discrete $Z_N$ symmetry as in our model. 

We reiterate and emphasize here that Eq.~\eqref{eq:0nu} is a model-independent sum rule for any hidden charges that the SM fermions may carry, and so provides a more model independent way of looking at the $0\nu\beta\beta$ experiments.

\section{\label{sec:Configs} Viable configurations}
In the lepton sector, it is a particular challenge for scenarios with Dirac neutrinos to naturally explain the bi-large mixing pattern observed in neutrinos, the lepton masses, and simultaneously suppress lepton flavour violations (LFVs) with a scale that is not high (of a few TeV order). We demonstrate in this section that all this is possible for Dirac neutrinos in the MCRS model with natural -- viz. anarchic -- Yukawa coupling, by finding configurations in the parameter space that satisfy all these requirements.

In our search, we scan through lepton mass matrices generated by varying lepton localization parameters ($c_L$, $c_E$ and $c_{\nu_R}$) and 5D Yukawa couplings ($\lambda_{5,ij}$) for those that could reproduce the observed lepton mass and mixing patterns while still satisfy LFV constraints. Since EWPT constraints generically set the KK scale at around 3~TeV, we conduct our search for $M_{KK} = 3$~TeV.

When generating the 5D Yukawa couplings, we take $|\lambda_{5,ij}| \in [0.5,2.0]$ so that they are perturbative, and no unnatural hierarchies would arise; we put no restrictions on the complex phases. Searching for the lepton localization parameters needs more guidelines. Electroweak constraints from the $Z\ra\tau\tau,\,\mu\mu$ branching ratios require $c_L > 1/2$ and $c_{E_3} < -1/2$~\cite{PR09}. Next, since we assume anarchic Yukawa couplings, mixing angles essentially depend on the ratio of lepton wavefunctions. The bi-large mixing pattern then indicates that LH charged leptons have similar wavefunctions, and thus similar $c_L$'s. Once the $c_L$'s are given, the range of the RH lepton localizations are then set by the lepton masses.

For charged leptons, in order to reproduce the mass pattern we require the eigenvalues of the charged lepton mass matrices to be within $1\sigma$ error of the values given in~\cite{XZZ08} at the 1~TeV scale. Constraints from LFVs also need to taken into account, and we require the our configurations be such that
$Br(\mu \ra 3 e) < 10^{-12}$ and $Br(\tau \ra l_1 l_2 \bar{l_3}) < 10^{-7}$ hold~\footnote{We find that $\mu$-$e$ conversion do not place further constraints on the viable mass matrices once that from $\mu \ra 3 e$ are satisfied. We do not consider constraints from $l \ra l' \gamma$ here as they are UV sensitive in scenarios with Higgs localized on branes~\cite{ABP06}.}.

For neutrinos, we require not only the masses, but also the PMNS mixing matrix be reproduced. Taking the usual parametrization~\cite{PDG08}, the PMNS matrix, $V_{PMNS} = (U^e_L)^\hc U^\nu_L$, can be written as
\begin{equation}
V_{PMNS} =
\begin{pmatrix}
1 &  0       & 0 \\
0 &  c_{23}  & s_{23} \\
0 & -s_{23}  & c_{23} \\
\end{pmatrix} \cdot
\begin{pmatrix}
 c_{13}                  & 0 & s_{13}e^{-i\delta_{CP}} \\
 0                       & 1 & 0 \\
-s_{13}e^{i\delta_{CP}}  & 0 & c_{13} \\
\end{pmatrix} \cdot
\begin{pmatrix}
 c_{12} & s_{12} & 0 \\
-s_{12} & c_{12} & 0 \\
 0      & 0      & 1 \\
\end{pmatrix} \,,
\end{equation}
where $c_{ij}$ ($s_{ij}$) denotes $\cos\theta_{ij}$ ($\sin\theta_{ij}$) with $\theta_{ij} \in [0,\,\pi/2]$, and $\delta_{CP} \in [0,\,2\pi)$. Note that for Dirac neutrinos, the Majorana phases can and have been absorbed into the neutrino states. When setting the range to search for the allowed neutrino parameters, we use the $3\sigma$ bound on the values of neutrino oscillation parameters derived from a global $3\nu$ analysis of the current data~\cite{GGM08}:
\begin{gather}
\Delta m_{21}^2 = 7.67^{+0.67}_{-0.61} \times 10^{-5}\,\mathrm{eV^2} \,, \qquad
\Delta m_{31}^2 = 2.46^{+0.47}_{-0.42} \times 10^{-3}\,\mathrm{eV^2} \,, \\
\theta_{12} = 34.5^{+4.8}_{-4.0} \,, \qquad
\theta_{23} = 42.3^{+11.3}_{-7.7} \,, \qquad
\theta_{13} = 0.0^{+12.9}_{-0.0} \,, \qquad
\delta_{CP} \in [0,\,2\pi] \,,
\end{gather}
where the mixing angles are given in degrees. It turns out that only normal hierarchy is viable in our search.

\begin{table}[ht]
\begin{ruledtabular}
\begin{tabular}{cccc}
Config. & $c_L$ & $c_E$ & $c_{\nu_R}$ \\
\hline
1 & $\{0.5876,\,0.5476,\,0.5001\}$ & $\{-0.7245,\,-0.5882,\,-0.5216\}$ & $\{-1.247,\,-1.223,\,-1.278\}$ \\
2 & $\{0.5880,\,0.5456,\,0.5014\}$ & $\{-0.7211,\,-0.5917,\,-0.5213\}$ & $\{-1.333,\,-1.246,\,-1.223\}$ \\
3 & $\{0.5865,\,0.5454,\,0.5006\}$ & $\{-0.7242,\,-0.5899,\,-0.5217\}$ & $\{-1.223,\,-1.355,\,-1.245\}$ \\
4 & $\{0.5877,\,0.5377,\,0.5006\}$ & $\{-0.7249,\,-0.5947,\,-0.5203\}$ & $\{-1.321,\,-1.250,\,-1.224\}$ \\
5 & $\{0.5830,\,0.5328,\,0.5018\}$ & $\{-0.7276,\,-0.6005,\,-0.5229\}$ & $\{-1.254,\,-1.224,\,-1.384\}$
\end{tabular}
\end{ruledtabular}
\caption{\label{tb:clep} Fermion localization parameters for the charged leptons and the RH neutrinos.}
\end{table}

\begin{table}[ht]
\begin{ruledtabular}
\begin{tabular}{ccccc}
Config. & Charged lepton masses (MeV) & Neutrino masses (meV) & $\delta_{CP}$ & $\{\theta_{12},\,\theta_{23},\,\theta_{13}\}$ ($^\circ$) \\
\hline
1 & $\{0.4959,\,104.7,\,1780\}$ & $\{1.4,\,8.9,\,50\}$   & $-0.47$ & $\{39,\,36,\,2.7\}$ \\
2 & $\{0.4959,\,104.7,\,1779\}$ & $\{0.22,\,8.5,\,47\}$  & $2.5$   & $\{32,\,42,\,6.6\}$ \\
3 & $\{0.4959,\,104.7,\,1779\}$ & $\{0.26,\,9.0,\,47\}$  & $1.3$   & $\{35,\,38,\,1.9\}$ \\
4 & $\{0.4959,\,104.7,\,1780\}$ & $\{0.13,\,8.7,\,47\}$  & $2.4$   & $\{35,\,53,\,9.7\}$ \\
5 & $\{0.4959,\,104.7,\,1780\}$ & $\{0.096,\,9.1,\,53\}$ & $1.5$   & $\{37,\,49,\,12\}$
\end{tabular}
\end{ruledtabular}
\caption{\label{tb:nup} Mass eigenvalues and mixing parameters of the viable configurations compatible with current lepton flavour constraints and neutrino oscillation data assuming normal hierarchy.}
\end{table}

We record here five representative viable configurations found in our search. In Table~\ref{tb:clep}, we display the lepton localization parameters for each of these configurations, and in Table~\ref{tb:nup} the mass eigenvalues and the PMNS parameters obtained. The actual charged lepton and neutrino mass matrices for each configuration are collected in Appendix~\ref{app:LepM}. Note that $\theta_{13}$ is generically nonzero in all the viable configurations we found.

\section{\label{sec:Pheno} Phenomenology}
In general, KK excitations of gauge bosons and fermions have masses of the order of a few TeV. Along with the suppressed coupling to SM fields, this makes them hard to produce and detect at the LHC (at least initially). However, as was pointed out in~\cite{CusRS}, KK fermions with $(-+)$ BCs can be very light in comparison. In particular, this means that the $SU(2)_R$ doublet partner of the RH electron, $\tilde{\nu}_R$, can be much lighter than all the other KK excitations in the spectrum.

\begin{figure}[htbp]
\centering
\includegraphics[width=3in]{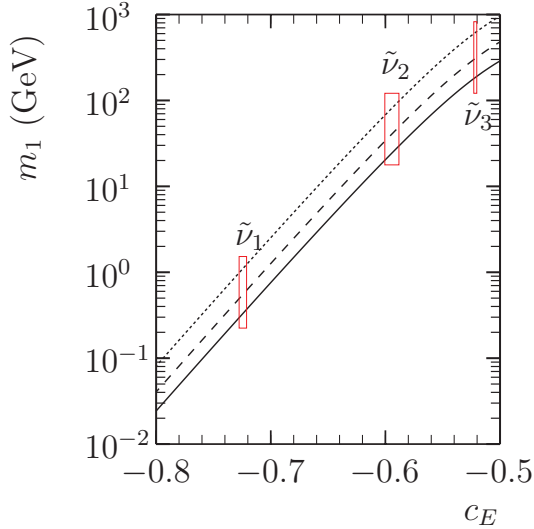}
\caption{\label{fig:KKFMass} The mass of first $(-+)$ KK fermion, $m_1$, as a function of the localization parameter, $c_E$. The width of the red boxes denotes for each generation of $\tilde{\nu}_R$, the range of $c_E$ variation in the five representative configurations listed in Table~\ref{tb:clep}. The solid, dashed and dotted lines denote cases where $M_{KK} = 3,\,5,\,10$~TeV.}
\end{figure}

The masses of the $(-+)$ KK $\tilde{\nu}_R$, $m_n$, are determined by their BCs:
\begin{equation}
\frac{J_{c_E+1/2}(m_n/k)}{Y_{c_E+1/2}(m_n/k)} =
\frac{J_{c_E-1/2}(m_n e^{k r_c\pi}/k)}{Y_{c_E-1/2}(m_n e^{k r_c\pi}/k)} \,.
\end{equation}
As was shown in~\cite{AS04}, when $c_E < -0.5$, the first $(-+)$ KK fermion becomes much lighter than the first $(++)$ KK fermion. We display its mass as a function of $c_E$ in Fig.~\ref{fig:KKFMass}.
We see that for the five representative configurations listed in Table~\ref{tb:clep}, we have an electron-like neutrino ($\tilde{\nu}_1$) with mass 175 - 222~MeV, a muon-like neutrino ($\tilde{\nu}_2$) with mass
16 - 24~GeV, and a tau-like neutrino ($\tilde{\nu}_3$) with mass 168 - 180~GeV. On the other hand, the KK excitations of $\nu_{L,R}$ are all heavier than 3~TeV as noted above. We note that the appearance of these relative light $(-+)$ KK neutrinos is a direct consequence of the localization of charged leptons we used to fit their masses.

Below, we will focus on the phenomenological consequences of these $(-+)$ KK neutrinos, which can have the greatest impact at the LHC.

\subsection{Effective couplings}
To proceed, we first work out the effective couplings of $(-+)$ KK neutrinos to SM fields. Especially interesting are the $W\tilde{\nu}_{iR}e_{iR}$ and $Z\bar{\tilde{\nu}}_{iR}\tilde{\nu}_{iR}$ couplings. The $Z\bar{\tilde{\nu}}_R\nu_R$ coupling is further suppressed by $m_\nu/M_W$, and so will be small. Similar couplings in the quark sector are also suppressed by the small effective 4D Yukawa couplings.

\begin{figure}[htbp]
\centering
\includegraphics[width=4in]{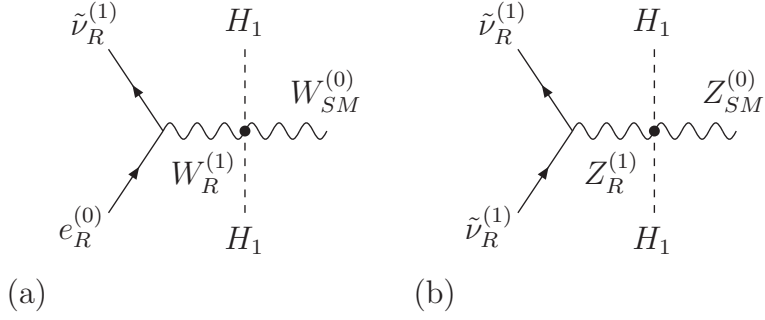}
\caption{\label{fig:SMmixing} Effective couplings of $\tilde{\nu}_R$ from gauge mixing.}
\end{figure}

\begin{figure}[htbp]
\centering
\includegraphics[width=4.5in]{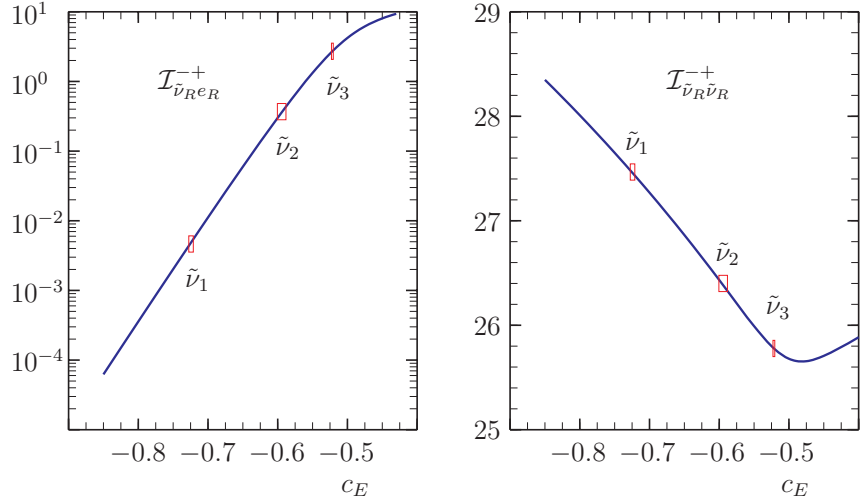}
\caption{\label{fig:EE_WZ} The overlapping function $\mathcal{I}^{-+}_{\tilde{\nu}_R e_R}$
and $\mathcal{I}^{-+}_{\tilde{\nu}_R\tilde{\nu}_R}$ vs $c_E$. The width of the red boxes denotes the range of $c_E$ variation in the five representative configurations.}
\end{figure}

The effective couplings of $\tilde{\nu}_R$ to SM $W$ and $Z$ arise primarily due to the mixing of gauge boson modes through interactions with the SM Higgs on the IR brane. Their leading contributions are depicted in Fig.~\ref{fig:SMmixing}~\cite{CNW08}.
We can parametrize them as
\begin{equation}
W\tilde{\nu}_{iR}e_{iR}:\;\frac{g_L}{\sqrt{2}}\,r_i \,, \qquad
Z\bar{\tilde{\nu}}_{iR}\tilde{\nu}_{iR}:\;
\frac{g_L}{\cos\theta_W}\gamma^\mu\!\left[z_{Li}\hat{L} + z_{Ri}\hat{R}\right] \,,
\end{equation}
where $\hat{L}$ and $\hat{R}$ are the usual chiral projectors, $\theta_W$ is the Weinberg angle, and
$g_L \equiv e/\sin\theta_W$. Since the gauge coupling of $SU(2)_L$ and $SU(2)_R$ are expected to be of the same order, we assume for simplicity that $g_L = g_R$. We can estimate~\footnote{We neglect here the effects of mixing between zero and KK modes, which has been argued to be small~\cite{EWGau07}.}:
\begin{equation}\label{eq:rnu}
r_i\sim\frac{g_L^2 v_1^2}{2 M^2_{-+}}\mathcal{I}^{-+}_{\tilde{\nu}_R e_R}(c_{E_i}) \,, \qquad
z_{Li} \sim z_{Ri} \sim \frac{g_L^2 v_1^2}{2M^2_{-+}}\mathcal{I}^{-+}_{\tilde{\nu}_R\tilde{\nu}_R}(c_{E_i})\,,
\end{equation}
where $M_{-+}$ denotes the mass of the first KK excitation of the $(-+)$ gauge boson. The quantities
$\mathcal{I}^{-+}_{ff'}$ are products of wavefunction overlap integrals from that between the fermions and the $(-+)$ KK gauge modes, and that between the $(-+)$ KK gauge modes and the SM gauge boson on the IR brane. We plot their dependence on $c_E$ in Fig.~\ref{fig:EE_WZ}.

Given the $c_E$'s from our representative configurations above, we have:
\begin{gather}
\{r_1,\,r_2,\,r_3\}\sim\{2.0 \times 10^{-3},\,0.15,\,1.0\} \times 10^{-3} \times \left(\frac{3\,\mrm{TeV}}{M_{-+}}\right)^2 \,, \\
\{z_{1},\,z_{2},\,z_{3}\}\sim\{9.7,\,9.3,\,9.1\}\times 10^{-3} \times \left(\frac{3\,\mrm{TeV}}{M_{-+}}\right)^2 \,,
\end{gather}
where $z_i$ denotes either $z_{Li}$ or $z_{Ri}$.

\subsection{Low energy tests}
Since $\tilde{\nu}_1$ is expected to be heavier than 170~MeV, only charged mesons heavier than the $\pi$ can
decay into a $\tilde{\nu}_1 e$. The most stringent limit comes from the decay of the charged kaon. The negative result from the search for additional peaks in the $e^+$ spectrum of the
$K^+\ra e^+ \tilde{\nu}_1$ decay sets a bound of $|r_1|^2 < 10^{-6}$ for a 160 to 220~MeV neutrino~\cite{PDG08}, and our estimate of $r_1$ above shows that it is well within this bound. The Fermi constant, $G_F$, best determined by the muon decay is thus not be modified at tree level by the
existence of a light $\tilde{\nu}_1$.

Another constraint comes from the measurement of the number of light neutrinos below the $Z$ pole. It is determined by measuring the invisible $Z$ decay at LEP~\cite{ZEWP}:
\begin{equation}
N_\nu=\frac{\Gamma_{inv}}{\Gamma_\nu^{SM}}=2.9840 \pm 0.0082\,.
\end{equation}
The width for $Z$ decays into $\tilde{\nu}_i$ pair is given by:
\begin{equation}
\Gamma(Z\ra\bar{\tilde{\nu}}_i\tilde{\nu}_i) =
\frac{G_F M_Z^3}{3\sqrt{2}\pi}\sqrt{1 - 4y_i}
\left[(z_{Li}^2 + z_{Ri}^2)(1 - y_i) + 6 y_i z_{Li}z_{Ri}\right] \,, \quad
y_i\equiv\frac{M^2_{\tilde{\nu}_i}}{M_Z^2} \,.
\end{equation}
Since $\tilde{\nu}_{2}$ can decay into charged final states immediately after being produced (see below), only $\tilde{\nu}_1$ can live to escape the detector without leaving any tracks. Therefore, the LEP limit requires that
\begin{equation}
z^2_{L1} + z^2_{R1} \leq 0.096 \; (95\%\,\mrm{CL}) \,,
\end{equation}
which is larger than our estimates above.

\subsection{Decays of $\tilde{\nu_i}$}
The $\tilde{\nu}_i$ KK neutrinos are unstable states, and their decay modes depend crucially on their masses.
For $\tilde{\nu}_3$, we expect it to decay predominantly into $\tau\,W$. The width is given by
\begin{equation}
\Gamma_{\tilde{\nu}_3} = \frac{g_L^2 r^2_3}{64\pi}\frac{M^3_{\tilde{\nu}_3}}{M_W^2}
(1 - w_3^{-1})^2(1 + 2w_3^{-1}) \,, \quad w_i\equiv\frac{M^2_{\tilde{\nu}_i}}{M_W^2} \,.
\end{equation}
For $M_{KK}=3$ TeV and $M_{\tilde{\nu}_3} = 175$~GeV, the width is about $1.5 \times 10^{-6}$~GeV.

For the (much) lighter $\tilde{\nu}_{1,2}$, three body decays are dominant. For
$m_{f,f'} \ll M_{\tilde{\nu}_i}$, the tree level differential width of the decay
$\tilde{\nu}_i \ra e_i\bar{f}f'$ takes the form:
\begin{equation}
\frac{d\Gamma}{d x_f} = N_c |V_{ff'}|^2 |r_i|^2 \frac{G_F^2
M^5_{\tilde{\nu}_i}}{16\pi^3} \frac{x_f^2(1 - x_f -
\eps_i)^2}{(1-x_f)- w_i x_f (1 - x_f - \eps_i)} \,, \quad
\eps_i\equiv\frac{m^2_{e_i}}{M^2_{\tilde{\nu}_i}} \,,
\end{equation}
where $x_f$ is the reduced energy of $\bar{f}$, $0 \leq x_f \leq 1-\epsilon_i$, and $V_{ff'}$ denotes the appropriate CKM mixing matrix element.

For $\tilde{\nu}_1$, the dominant decay channel is the charged current (CC) decay
$\tilde{\nu}_1\ra e e^+ \nu_e$:
\begin{equation}
\Gamma^\mrm{CC}_1 \sim |r_{1}|^2\frac{G_F^2 M^5_{\tilde{\nu}_1}}{192 \pi^3} = 0.73 \times 10^{-17}
\times |r_{1}|^2 \times \left(\frac{M_{\tilde{\nu}_1}}{200\,\mrm{MeV}}\right)^5\mrm{GeV} \,.
\end{equation}
Due to phase space suppression, we ignore the small contribution of $\tilde{\nu}_1\ra e \mu^+ \nu_\mu$. The
$e\,\pi$ mode is also negligible, while virtual $Z$ mediated amplitudes are unimportant. The lifetime of $\tilde{\nu}_1$ is then estimated to be
\begin{equation}
\tau_{\tilde{\nu}_1} \sim 2.3 \times 10^{4} \times \left(\frac{M_{KK}}{3\,\mrm{TeV}}\right)^4
\times \left(\frac{200\,\mrm{MeV}}{M_{\tilde{\nu}_1}}\right)^5\mrm{sec} \,.
\end{equation}

For $\tilde{\nu}_2$, the main CC decays channels are
$\tilde{\nu}_2\ra\mu\bar{l}\nu_l,\,\mu \bar{d} u,\,\mu\bar{s}c$. Since $M_{\tilde{\nu}_2}\sim\mathcal{O}(10)$~GeV, the final state fermion masses can be ignored, and the CC decay width is given by
\begin{equation}
\Gamma(\tilde{\nu}_2\ra\mu\bar{f}f') = N_c |V_{ff'}|^2 |r_{2}|^2 \frac{G_F^2 M_W^5}{16\pi^3}h(w_2) \,,
\end{equation}
where
\begin{equation}
h(x)= \frac{6x - 3x^2 - x^3 + (1 - x)\ln(1-x)}{6x\sqrt{x}} \,.
\end{equation}
For $M_{\tilde{\nu}_2} \in [16,\,24]$~GeV, $h(w_2) \in [4.1,\,2.7]$ is almost a liner function in $M_{\tilde{\nu}_2}$. The total CC decay width of a $20$~GeV $\tilde{\nu}_2$ is estimated to be
\begin{equation}
\Gamma_2^\mrm{CC} \sim |r_{2}|^2 \frac{G_F^2 M_W^5}{16\pi^3}h(w_2) \times [1+1+1+3+3]
\sim 0.027|r_{2}|^2\,\mrm{GeV} \,,
\end{equation}
and so its lifetime is $\sim 1.2 \times 10^{-15}$ sec for $M_{KK} = 3$~TeV.

\subsection{Production of $\tilde{\nu_i}$ at the LHC}
Since $\tilde{\nu}_1$ is much lighter than a GeV, it is not expected to be seen at the LHC due to the large background. We thus focus on $\tilde{\nu}_{2,3}$, which are heavy when compared to the GeV scale. For $\tilde{\nu}_2$, it can be detected via the process $u\bar{d}\ra\tilde{\nu}_2\mu^+\ra\mu^+\mu^-e(\tau)\bar{\nu}$.  The final state will involve apparent lepton flavor violation plus missing energy with the $\mu^+\mu^-$ pair not in resonance. These are
characteristic heavy neutrino signatures. Similarly, $\tilde{\nu_3}$ can be detected via the process $u\bar{d}\ra\tau^+\tilde{\nu}_3\ra\tau^+\tau^-W$, where a $W$ jet plus $\tau$ jets are expected and the $\tau$ jets are not in resonance.

The tree-level cross-section for $u\bar{d} \ra W^+ \ra \tilde{\nu}_i e_i^+$ at the parton level can be  straightforwardly worked out:
\begin{equation}
\hat{\sigma}(\hat{s}) = \frac{1}{N_c}\frac{g_L^4|V_{ud}|^2 r^2_i}{192\pi}\,\frac{1}{\hat{s}}\,
\frac{(1 - x_{N_i})^2}{(1 - x_W)^2}\left(1 + \frac{x_{N_i}}{2}\right) \,, \qquad
x_{N_i}\equiv\frac{M_{N_i}^2}{\hat{s}} \,, \quad x_W\equiv\frac{M_{W}^2}{\hat{s}} \,,
\end{equation}
where $\hat{s}$ is the center-of-mass (CM) energy of the two colliding partons. The production cross-section at LHC is then obtained from the convolution with parton distribution functions (PDFs):
\begin{equation}
\sigma(pp\ra \tilde{\nu}_i e_i^+) =
\int\!dx_1 dx_2\,2f_u(x_1)f_d(x_2)\hat{\sigma}(x_1 x_2 s)\theta(1-x_{N_i}) \,, \quad
s \equiv \frac{\hat{s}}{x_1 x_2} \,.
\end{equation}

\begin{figure}[htbp]
\centering
\includegraphics[width=3.5in]{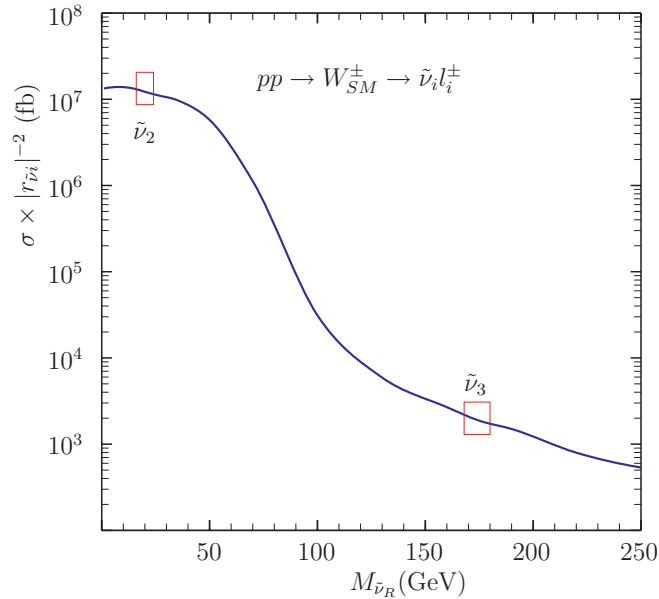}
\caption{\label{fig:LHC_lN} The associated $\tilde{\nu}_R$ production cross section as a function of its mass at the LHC with CM energy $\sqrt{s}=14$ TeV. The width of the red boxes denotes the range of $c_E$ variation in the five representative configurations.}
\end{figure}
We plot in Fig.~\ref{fig:LHC_lN} the total production cross-section as a function of the mass of $\tilde{\nu}$. The total production includes both $\bar{\tilde{\nu}}_i e_i^\pm$ productions, and we have used the MSTW 2008 PDFs~\cite{MSTW}. From it, one can estimate the total single $\tilde{\nu}_R$ production cross section at $\sqrt{s}=14$~TeV to be $\sim 0.3$~fb and $\sim 10^{-3}$~fb for $\tilde{\nu}_2$ and $\tilde{\nu}_3$ respectively.

\section{\label{sec:Conc} Conclusions}
We have shown that Dirac neutrinos can be naturally implemented in a MCRS setting with
$SU(2)_L\times SU(2)_R \times U(1)_{B-L}\times U(1)_X$ bulk gauge symmetry via the KW mechanism, which leaves
a minimal gauged $Z_3$ after the $U(1)_X$ is broken on the UV brane. We have seen in the case of normal neutrino hierarchy, lepton masses and mixing patterns can be successfully reproduced with just the RS
anarchic 5D flavour structure; no symmetries or otherwise need be imposed. Furthermore, this can all be done with a 3~TeV level KK scale as set by EWPTs, while also keeping LFVs under control. However, when the neutrinos have an inverted hierarchy or are degenerate, excessive fine tuning in the Yukawa couplings that plagues the usual 4D Dirac neutrino scenarios is very difficult to avoid without pushing the KK scale much too high to be relevant.

With neutrinos in the normal hierarchy, the viable lepton configurations we found generically predicts a nonzero $\theta_{13}$. Moreover, neither are small values of $\theta_{13}$ favored in particular. Thus, a measured value of $\theta_{13}$ that is very close to zero will make our Dirac neutrino scenario unlikely.

Another interesting feature of our scenario is the existence of light $(-+)$ RH KK neutrinos, $\tilde{\nu}_i$, that are $SU(2)_R$ partners to the RH charged leptons. Their masses are sensitive to their localization in the 5D bulk, viz. the bulk mass parameters $c_{E_i}$ of the RH lepton doublets $E_i$, which are determined by the charged lepton and neutrino data. For the viable lepton configurations we found, we have
$M_{\tilde{\nu}_1} \sim 170 $~MeV, $M_{\tilde{\nu}_2} \sim 20$~GeV, and $M_{\tilde{\nu}_3} \sim 180$~GeV, all much lighter than the 3~TeV level first KK gauge bosons. For the LHC, $\tilde{\nu}_2$ will be the most interesting as it has a large enough production cross-section for the search to be worthwhile. Of the other two generations, $\tilde{\nu}_1$ is far too light -- although it is worth noting that it has a very long lifetime of $\sim 10^4$~s -- while $\tilde{\nu}_3$ has too small a production rate at the LHC. It is intriguing to note that while it will be very difficult to find these KK neutrinos, if found their mass pattern may serve as a way to measure the localization parameters $c_{E_i}$.

\begin{acknowledgments}
The research of W.F.C. is supported by the Taiwan NSC under Grant No. 96-2112-M-007-020-MY3. The research of J.N.N. is partially supported by the Natural Science and Engineering Council of Canada. The research of J.M.S.Wu is supported in part by the Innovations und Kooperationsprojekt C-13 of the Schweizerische Universitaetskonferenz SUK/CRUS.
\end{acknowledgments}

\appendix

\section{\label{app:4DDN} Dirac neutrinos from $SU(2)_L \times U(1)_Y \times Z_N$ in 4D}
We review here how Dirac neutrinos are commonly implemented in 4D gauge theories with a $Z_N$ extension to the SM gauge group. For simplicity, we consider here the case of just one generation of fermions species:
$Q = (u,d)_L$, $u_R$, $d_R$, $L=(\nu,e)_L$, $e_R$, and a RH neutrino $n_R$. Generalization to three generations is straightforward, and does not alter the physics.

Under the $Z_N$ symmetry, fermion fields transform as
\begin{equation}
\psi_f \ra e^{\frac{2\pi i\alpha}{N}}\psi_f \,, \quad  \alpha = 0,1,\dots,N-1 \,,
\end{equation}
where $\alpha$ is the discrete charge of $\psi_f$. For $Z_N$ to remain unbroken, the SM Higgs is required to be a singlet under the $Z_N$, i.e. $\alpha_H = 0$. To have quark and charged lepton masses, we have the following constraints:
\begin{equation}
\alpha_Q-\alpha_u = 0 \bmod N \,, \quad
\alpha_Q-\alpha_d = 0 \bmod N \,, \quad
\alpha_L-\alpha_e = 0 \bmod N \,.
\end{equation}
To have a massive Dirac neutrino, we further require
\begin{equation}
\alpha_L-\alpha_n = 0 \bmod N \,, \quad 2\alpha_n \neq N \,.
\end{equation}
This constraint forbids a Majorana mass term in the neutrino mass matrix and immediately rules out $Z_2$ as a viable discrete symmetry. We see that the dimension five operator, $LLHH$, is automatically forbidden.

There are many solutions that satisfy all the constraints above. One such solution is
\begin{equation}
\label{eq:charges}
\alpha_L = \alpha_e = \alpha_n = 1 \,, \quad \alpha_Q = \alpha_u = \alpha_d = 2 \,.
\end{equation}
In this case, the discrete group is $Z_3$, which is also the smallest. Interestingly, gauge invariant dimension six operators
\begin{equation}
\ovl{d^c}u\ovl{Q^c}L \,, \; \ovl{Q^c}Q \ovl{u^c}e \,, \; \ovl{Q^c}Q \ovl{Q^c}L \,, \;
\ovl{d^c}u\ovl{u^c}e \,, \; \ovl{u^c}u\ovl{d^c}e \,, \; uddn \,,
\end{equation}
are all forbidden. The superscript above denotes charge conjugation, and for simplicity we have dropped the fermion chirality label. On the other hand, dimension nine operators which cause neutron anti-neutron oscillations are allowed by the $Z_3$. However, this can be suppressed by have a $Z_4$ symmetry instead.

\section{\label{app:LepM} Lepton mass matrices}
In this appendix, we give the mass matrices for the charged leptons and the neutrinos, $M_e$ and $M_\nu$, of the five viable configurations presented in Table~\ref{tb:nup}. All mass matrices are given in units of GeV.
\begin{itemize}
\item Configuration~1: 
\begin{align}
M_e &=
\begin{pmatrix}
 0.3860 + 0.2173\,i & 25.64 + 2.695\,i & 15.86 + 145.2\,i \\
-1.093  + 0.9150\,i & 139.4 - 7.216\,i & 137.3 + 427.4\,i \\
 2.592  - 1.045\,i  & 10.75 - 122.8\,i & 1709  + 0.4385\,i
\end{pmatrix} \times 10^{-3} \\
M_{\nu} &=
\begin{pmatrix}
 4.084 - 0.7790\,i  & 7.347 + 5.348\,i & -0.4597 + 0.9361\,i \\
-4.470 - 0.09735\,i & 30.73 + 9.532\,i &  2.343  + 1.954\,i \\
-12.76 + 3.118\,i   & 33.63 + 7.340\,i &  7.523  + 0.3198\,i
\end{pmatrix} \times 10^{-12}
\end{align}
\item Configuration~2: 
\begin{align}
M_e &=
\begin{pmatrix}
 0.4000 - 0.04051\,i & -1.217 - 22.11\,i & -160.7 - 53.23\,i \\
-0.2626 - 1.041\,i   &  62.33 - 54.21\,i & -402.2 - 107.4\,i \\
 0.5107 - 2.120\,i   &  237.6 + 165.0\,i &  1698  + 3.339\,i
\end{pmatrix} \times 10^{-3} \\
M_{\nu} &=
\begin{pmatrix}
0.2051 - 0.05470\,i & 3.496 - 1.663\,i  & -5.386 - 3.753\,i \\
0.3466 - 0.1070\,i  & 15.26 - 0.6531\,i & 17.65  - 1.742\,i \\
0.6527 - 0.1867\,i  & 15.37 - 1.692\,i  & 37.27  - 1.687\,i
\end{pmatrix} \times 10^{-12}
\end{align}
\item Configuration~3:
\begin{align}
M_e &=
\begin{pmatrix}
0.3280 - 0.01371\,i & -19.12 + 0.7429\,i & -107.2 - 47.15\,i \\
0.7746 + 0.4967\,i  &  136.1 + 34.55\,i  &  318.3 + 388.3\,i \\
1.868  - 0.6988\,i  &  147.1 + 3.573\,i  &  1693  - 2.069\,i
\end{pmatrix} \times 10^{-3} \\
M_{\nu} &=
\begin{pmatrix}
 1.544 + 4.049\,i & -0.03519 + 0.01279\,i & -0.5576 - 4.935\,i \\
 3.637 - 32.64\,i &  0.2820  - 0.05918\,i & 14.32   - 1.458\,i \\
-1.294 - 27.29\,i & -0.2914  + 0.08477\,i & 12.44   + 7.564\,i
\end{pmatrix} \times 10^{-12}
\end{align}
\item Configuration~4: 
\begin{align}
M_e &=
\begin{pmatrix}
0.1754 - 0.1356\,i & 33.33 + 0.7621\,i & -102.6 + 208.7\,i \\
0.3241 - 0.7123\,i & 77.43 + 83.87\,i  & -598.0 - 144.9\,i \\
0.6124 + 2.938\,i  & 36.08 - 237.3\,i  &  1635  - 5.170\,i
\end{pmatrix} \times 10^{-3} \\
M_{\nu} &=
\begin{pmatrix}
-0.06034 - 0.1162\,i & 1.685 - 4.426\,i & -7.721 - 5.338\,i \\
 0.8883  - 0.6099\,i & 17.03 - 1.050\,i & 18.10  - 0.4207\,i \\
 1.117   - 0.9943\,i & 16.61 - 2.132\,i & 36.01  - 2.739\,i
\end{pmatrix} \times 10^{-12}
\end{align}
\item Configuration~5: 
\begin{align}
M_e &=
\begin{pmatrix}
0.5140 - 0.002409\,i & 15.71 - 5.568\,i &  26.75 + 150.2\,i \\
1.363  + 2.170\,i    & 89.12 + 68.98\,i & -668.1 - 0.5164\,i \\
1.252  + 2.437\,i    & 59.13 - 168.0\,i &  1632  - 4.105\,i
\end{pmatrix} \times 10^{-3} \\
M_{\nu} &=
\begin{pmatrix}
1.985 - 2.990\,i & 0.05532 - 10.88\,i & -0.02715 + 0.01001\,i \\
10.32 - 7.875\,i & 19.45   - 1.781\,i & -0.1622  + 0.004459\,i \\
6.000 - 15.18\,i & 44.22   + 1.591\,i &  0.1246  + 0.004404\,i
\end{pmatrix} \times 10^{-12}
\end{align}
\end{itemize}


\begin{thebibliography}{99}
\bibitem{KW88}
L.~M.~Krauss and F.~Wilczek,
%``Discrete Gauge Symmetry in Continuum Theories,''
Phys. Rev. Lett.  {\bf 62}, 1221 (1989).

\bibitem{B89}
T.~Banks,
%``EFFECTIVE LAGRANGIAN DESCRIPTION OF DISCRETE GAUGE SYMMETRIES,''
Nucl. Phys. B {\bf 323}, 90 (1989).

\bibitem{PK90}
J.~Preskill and L.~M.~Krauss,
%``Local Discrete Symmetry And Quantum Mechanical Hair,''
Nucl. Phys. B {\bf 341}, 50 (1990).

\bibitem{RS}
L.~Randall and R.~Sundrum,
%``A large mass hierarchy from a small extra dimension,''
Phys. Rev. Lett. {\bf 83}, 3370 (1999). %[arXiv:hep-ph/9905221].

\bibitem{Gher03}
T.~Gherghetta,
%``Dirac neutrino masses with Planck scale lepton number violation,''
Phys. Rev. Lett.  {\bf 92}, 161601 (2004). %[arXiv:hep-ph/0312392].

\bibitem{APS05}
K.~Agashe, G.~Perez and A.~Soni,
%``Flavor structure of warped extra dimension models,''
Phys. Rev. D {\bf 71}, 016002 (2005). %[arXiv:hep-ph/0408134].

\bibitem{FermLoc}
N.~Arkani-Hamed and M.~Schmaltz,
%``Hierarchies without symmetries from extra dimensions,''
Phys. Rev. D {\bf 61}, 033005 (2000), %[arXiv:hep-ph/9903417].
Y.~Grossman and M.~Neubert,
%``Neutrino masses and mixings in non-factorizable geometry,''
Phys. Lett. B {\bf 474}, 361 (2000), %[arXiv:hep-ph/9912408].
T.~Gherghetta and A.~Pomarol,
%``A warped supersymmetric standard model,''
Nucl. Phys. B {\bf 602}, 3 (2001). %[arXiv:hep-ph/0012378].

\bibitem{HS01}
S.~J.~Huber and Q.~Shafi,
%``Fermion Masses, Mixings and Proton Decay in a Randall-Sundrum Model,''
Phys. Lett. B {\bf 498}, 256 (2001). %[arXiv:hep-ph/0010195].

\bibitem{CNW08}
W.~F.~Chang, J.~N.~Ng and J.~M.~S.~Wu,
%``Testing Realistic Quark Mass Matrices in the Custodial Randall-Sundrum Model
%with Flavor Changing Top Decays,''
Phys. Rev. D {\bf 78}, 096003 (2008); %arXiv:0806.0667 [hep-ph].
%``Flavour Changing Neutral Current Constraints from Kaluza-Klein Gluons and
%Quark Mass Matrices in RS1,''
Phys. Rev. D {\bf 79}, 056007 (2009). %arXiv:0809.1390 [hep-ph].

\bibitem{CusRS}
K.~Agashe, A.~Degado, M.~J.~May and R.~Sundrum,
%``RS1, custodial isospin and precision tests,''
JHEP {\bf 0308}, 050 (2003)

\bibitem{ADS03}
K.~Agashe, A.~Delgado and R.~Sundrum,
%``Grand unification in RS1,''
Annals Phys. {\bf 304}, 145 (2003). %[arXiv:hep-ph/0212028].

\bibitem{CFA08}
C.~Csaki, A.~Falkowski and A.~Weiler,
%``The Flavor of the Composite Pseudo-Goldstone Higgs,''
JHEP {\bf 0809}, 008 (2008). %[arXiv:0804.1954 [hep-ph]].

\bibitem{CFA09}
C.~Csaki, A.~Falkowski and A.~Weiler,
%``A Simple Flavor Protection for RS,''
Phys. Rev. D {\bf 80}, 016001 (2009). %[arXiv:0806.3757 [hep-ph]].

\bibitem{CDGG08}
C.~Csaki, C.~Delaunay, C.~Grojean and Y.~Grossman,
%``A Model of Lepton Masses from a Warped Extra Dimension,''
JHEP {\bf 0810}, 055 (2008). %[arXiv:0806.0356 [hep-ph]].

\bibitem{PR09}
G.~Perez and L.~Randall,
%``Natural Neutrino Masses and Mixings from Warped Geometry,''
JHEP {\bf 0901}, 077 (2009). %[arXiv:0805.4652 [hep-ph]].

\bibitem{XZZ08}
Z.~Z.~Xing, H.~Zhang and S.~Zhou,
%``Updated Values of Running Quark and Lepton Masses,''
Phys. Rev. D {\bf 77}, 113016 (2008). %[arXiv:0712.1419 [hep-ph]].

\bibitem{ABP06}
K.~Agashe, A.~E.~Blechman and F.~Petriello,
%``Probing the Randall-Sundrum geometric origin of flavor with lepton flavor
%violation,''
Phys. Rev. D {\bf 74}, 053011 (2006). %[arXiv:hep-ph/0606021].

\bibitem{PDG08}
C. Amsler et al., Phys. Lett. B {\bf 667} (2008) 1.

\bibitem{GGM08}
M.~C.~Gonzalez-Garcia and M.~Maltoni,
%``Phenomenology with Massive Neutrinos,''
Phys. Rept. {\bf 460}, 1 (2008). %[arXiv:0704.1800 [hep-ph]].

\bibitem{AS04}
K.~Agashe and G.~Servant,
%``Warped unification, proton stability and dark matter,''
Phys. Rev. Lett. {\bf 93}, 231805 (2004). %[arXiv:hep-ph/0403143].

\bibitem{EWGau07}
K.~Agashe {\it et al.},
%``LHC Signals for Warped Electroweak Neutral Gauge Bosons,''
Phys. Rev. D {\bf 76}, 115015 (2007). %[arXiv:0709.0007 [hep-ph]].

\bibitem{ZEWP}
ALEPH Collaboration {\it et al.},
%``Precision electroweak measurements on the $Z$ resonance,''
Phys. Rept. {\bf 427}, 257 (2006).  %[arXiv:hep-ex/0509008].

\bibitem{MSTW}
A.~D.~Martin, W.~J.~Stirling, R.~S.~Thorne and G.~Watt,
%``Parton distributions for the LHC,''
arXiv:0901.0002 [hep-ph].
\end{thebibliography}
\end{document}